\documentclass[journal]{IEEEtran}
\usepackage{amsmath}
\usepackage{amssymb}
\usepackage{subfigure}
\usepackage[dvips]{graphicx}
\usepackage{cite}
\usepackage{flushend}
\usepackage{amsthm}
\newcommand{\argmin}{\operatornamewithlimits{argmin}}
\newcommand{\argmax}{\operatornamewithlimits{argmax}}
\usepackage{color,soul}

\title{Free Space Optical Communication with Spatial Modulation and Coherent Detection over H-K Atmospheric Turbulence Channels}
\author{Kostas~P.~Peppas,~\IEEEmembership{Senior Member,~IEEE} and P.~Takis~Mathiopoulos,~\IEEEmembership{Senior Member,~IEEE}
\thanks{K. P. Peppas is the Department of Informatics and Telecommunications, University of Peloponnese, 22100 Tripoli, Greece. He is also with with the Institute of Informatics and Telecommunications, National Centre for Scientific Research--``Demokritos," Patriarhou Grigoriou and Neapoleos, 15310 Agia Paraskevi, Athens, Greece (e-mail: kpeppas@iit.demokritos.gr)}
\thanks{P. T. Mathiopoulos is with the Department of Informatics and Telecommunications, National and Kapodistrian University of Athens, 15784 Zografou, Athens, Greece (e-mail: mathio@di.uoa.gr).}
}

\begin{document}

%\markboth{IEEE Journal on Selected Areas in Communications,~Vol.~X,
%No.~XX,~XXXXX~2014}{K. P. Peppas \MakeLowercase{\text \it{et al.}}: Free Space Optical Communication with Spatial Modulation and Coherent Detection over Atmospheric Turbulence Channels}

\maketitle
\begin{abstract}
The use of optical spatial modulation (OSM), which has been recently emerged as a power and bandwidth efficient pulsed modulation technique for indoor optical wireless communication, is proposed as a simple, low-complexity means of achieving spatial diversity in coherent free space optical (FSO) communication systems. In doing so, this paper makes several novel contributions as follows. It presents a generic analytical framework for obtaining
the Average Bit Error Probability (ABEP) of uncoded OSM with coherent detection in the presence of turbulence-induced fading. Although the framework is general enough to accommodate any type of models based on turbulence scattering, the focus in this paper is the H-K distribution. Although this distribution represents a very general scattering model valid over a wide range of atmospheric conditions, it is has not been considered in the past in conjunction with FSO systems possibly because of its mathematical complexity. The proposed analytical framework yields exact performance evaluation results for MIMO systems with two transmit– and an arbitrary number of receive apertures.
In addition, tight upper bounds are derived for the error probability for OSM systems
with an arbitrary number of transmit apertures as well as for convolutionally encoded signals.
The performance of OSM is compared to that of well established coherent FSO schemes,
employing spatial diversity at the transmitter or the receiver only. Specifically, it is shown that OSM can offer comparable performance with conventional coherent FSO schemes while
outperforming the latter in terms of spectral efficiency and hardware complexity.
Various numerical performance evaluation results are also presented and compared with equivalent results obtained by Monte Carlo simulations which verify the accuracy of the derived analytical expressions.
\end{abstract}
\begin{table*}[!t]
\hrulefill \vspace*{1pt}
\renewcommand{\arraystretch}{1.5}
\caption{List of Mathematical Notations}
\centering
\label{Tab:Notations}
\begin{tabular}{l}
\hline
\hline
$\jmath^2 = -1$ denotes the imaginary unit \\
$|z|$ denotes the magnitude of the complex number $z$\\
$\Re\{z\}$ denotes the real part of the complex number $z$\\
$\Im\{z\}$ denotes the real part of the complex number $z$\\
$f(x)=o[g(x)]$ as $x\rightarrow x_0$ if $\lim_{x\rightarrow x_0} \frac{f(x)}{g(x)} = 0$ \\
$\parallel \cdot \parallel_F^2$ denotes the square Frobenius norm \\
$(\cdot)^T$ denotes the matrix transpose \\
$\ast$ denotes convolution \\
${\mathbb E}\langle\cdot\rangle$  denotes expectation \\
$f_X(\cdot)$   denotes the Probability Density Function (PDF) of the random variable $X$  \\
$F_X(\cdot)$   denotes the Cumulative Distribution Function (CDF) of the random variable $X$ \\
$\mathcal{M}_X(\cdot)$   denotes the Moment Generating Function (MGF) of the random variable $X$ \\
$I_{a}\left(\cdot\right)$  is the modified Bessel function of the first kind and order $a$ \cite[eq. (8.431)]{B:Gradshteyn_00}\\
$K_{a}(\cdot)$ is the modified Bessel function of the second kind and order $a$ \cite[eq. (8.432)]{B:Gradshteyn_00} \\
$\Gamma\left(x\right) = \int_0^{\infty}\exp(-t)t^{x-1}\mathrm{d}t$  is the Gamma function \cite[eq. (8.310/1)]{B:Gradshteyn_00} \\
%$\Gamma\left(a, x\right) = \int_x^{\infty}\exp(-t)t^{a-1}\mathrm{d}t$  is the upper incomplete Gamma function \cite[eq. (8.350/2)]{B:Gradshteyn_00} \\
%$\gamma\left(a, x\right) = \int_0^{x}\exp(-t)t^{a-1}\mathrm{d}t$  is the lower incomplete Gamma function \cite[eq. (8.350/1)]{B:Gradshteyn_00} \\
%$G \,\substack{ m , n\\ p , q}\left[\cdot\right]$ is the Meijer's G-function \cite[eq. (8.2.1)]{B:Prudnikov3} \\
$Q(x) = \frac{1}{\sqrt{2\pi}}\int_x^{\infty}\exp(-t^2/2)\mathrm{d}t$ is the Gauss Q-function  \\
$W_{p,q}(\cdot)$ is the Whittaker function \cite[eq. (9.220)]{B:Gradshteyn_00} \\
$\mathrm{Pr}\{\cdot\}$ denotes the probability operator \\
$\hat{\cdot}$ denotes estimated value at the receiver side \\
 \hline
\end{tabular}
\end{table*}
\section{Introduction}\label{Sec:Intro}
Free-space optical (FSO) communication systems have
recently attracted great attention within the research community as well as for
commercial use. FSO systems can provide ultra-high data rates (at the order of multiple
gigabits per second), immunity to electromagnetic
interference, excellent security and large unlicensed bandwidth i.e. hundred and
thousand times higher than radio-frequency (RF) systems, along with low installation and operational cost \cite{J:Zhu1}.
%Because of their attractive features, FSO communication systems are today used for a wide variety
%of applications including "last mile" access, back-haul for
%wireless cellular networks, fiber backup, HDTV transmission,
%disaster recovery and rescue operations \cite{J:A_Majumdar}.

The challenge in employing such systems is that FSO links are highly vulnerable
due to the detrimental effects of attenuation under adverse weather conditions (e.g. fog),
pointing errors and atmospheric turbulence.
%Turbulence induced fading, also known as scintillation
%in optical communication terminology, results in time-varying fluctuations in the irradiance of the received optical
%laser beam in a similar fashion as fading in RF systems \cite{J:ghaspo1}.
One method to improve the reliability of the FSO link is to employ
spatial diversity, i.e. multiple-lasers and multiple-apertures to create
a multiple-input multiple-output (MIMO) optical channel.
Because of its low complexity, spatial diversity is a particularly attractive fading mitigation technique and performance enhancements have been extensively studied in many past research works in the field of FSO communications \cite{J:Garcia-ZambranaA, J:Navidpour, J:Peppas2, J:Peppas2012}.

In order to evaluate the impact of atmospheric turbulence on the performance of OSM, accurate models for the fading distribution
are necessary. For example the lognormal distribution is often used to
model weak turbulence conditions whereas the negative exponential and the K-distribution are used to model strong turbulence conditions \cite{J:Andrews}.
Other more general statistical models have also been proposed to model scintillation over all turbulence conditions,
including the Gamma-Gamma \cite{J:Al-Habash}, the lognormal-Rice (or Beckmann) \cite{J:Churnside}
the homodyned K distribution (H-K) \cite{J:Jakeman2} and the I-K \cite{J:AndrewsIK2,J:AndrewsIK3, J:PeppasIK} distributions.
All these three models are based on the argument that scintillation is a doubly stochastic random process modelling both small and large scale turbulence
effects. Besides, they agree well with measurement data and simulations for a wide range
of turbulence conditions.

In this paper, the H-K distribution is adopted to model turbulence-induced fading. The main reason for this choice is the fact that this distribution is based on a very general scattering model which is valid for a wide range of atmospheric conditions. It is also noted that the H-K distribution generalizes existing models such as the K-distribution.
The H-K distribution models the field of the optical wave as the sum of a deterministic component
and a random component, the intensity of which follows the Rice (Nakagami-$n$) distribution. The average intensity of the random
portion of the field is treated as a fluctuating quantity \cite{J:AndrewsIK2}.
It is important to underline that, to the best of our knowledge, in the open technical literature there have been no papers published analyzing and evaluating the performance of FSO systems over such channels,
because of the complicated mathematical form of their respective probability density functions (PDF).

Depending on their detection, FSO systems can be classified into two main categories, namely
coherent (heterodyne detection) and non-coherent (direct detection) systems.
Coherent FSO systems have the information bits encoded directly onto the electric field of the
optical beam. At the receiver, a local oscillator (LO) is employed
to extract the information encoded on the optical carrier electric field.
On the one hand, coherent FSO systems can provide significant performance enhancements
due to spatial temporal selectivity and heterodyne gain in comparison to direct detection systems.
Moreover, they are more versatile as
any kind of amplitude, frequency, or phase modulation can
be employed. On the other hand, coherent receivers are more difficult to implement as the
LO field should be spatially and temporally coherent with
the received field.
%The performance of coherent FSO systems has been addressed in several research, including \cite{J:Kiasaleh, J:Niu2, J:Aghajanzadeh, J:Niu3, J:Bayaki2} and references therein.

Recently, the so-called optical spatial modulation (OSM) has emerged as a power- and bandwidth-efficient
single-carrier transmission technique for optical wireless communication systems \cite{J:OSM2011, J:OSMCOML2011, J:FathHaas}.
This spatial diversity scheme, initially proposed in \cite{C:ChauandYuSpatial} and further
investigated in \cite{J:Mesleh, J:TrellisCodedSM}, employs a simple modulation
mechanism that foresees to activate just one out of several MIMO transmitters at any time instant and to use the index of the activated transmitter
as an additional dimension for conveying implicit information.
It has been shown that OSM can increase the data rate by
base two logarithm of the number of transmit units \cite{J:OSM2011}. Also,
OSM can increase the data rate by by a factor of 2 and 4, respectively, as compared to on-off keying (OOK)
and pulse position modulation (PPM) \cite{J:OSM2011, J:OSMCOML2011}.
It is underlined that such performance gains are obtained
with a significant reduction in receiver complexity and system
design. %Furthermore, no synchronization is necessary among MIMO elements and
%low-complexity non-coherent detection may be used instead
%of multiple-stream detection, usually employed in conventional MIMO systems \cite{J:RenzoProceedingsIEEE}.
%In general, OSM seems to be a promising candidate technology for optical MIMO
%systems that exploit the so-called "massive multi-functional MIMO (MF-MIMO)" paradigm \cite{J:HanzoProcIEEE}.
%For such systems, improved performance and energy efficiency can
%be achieved by employing a large number of
%transmitters to illuminate the target coverage area
%as well as receiver arrays consisting of hundreds of detectors.

Because of the above mentioned advantages of OSM over other more conventional transmission schemes and given
the wide applicability of FSO, it is of interest to investigate the potential performance enhancements obtained by incorporating OSM in FSO systems.
However, in general this research topic has not been dealt within our research community. Only recently, there have been papers published in the open technical literature dealing with performance analysis
studies of FSO systems employing spatial modulation and operating in the presence of atmospheric turbulence,
e.g. see \cite{J:Hwang} and \cite{J:Ozbilgin}.
Specifically, in \cite{J:Hwang}, the combination of subcarrier intensity modulation and spatial modulation with receiver diversity was proposed to enhance the performance of
intensity modulated direct detection (IM/DD) FSO systems.
In \cite{J:Ozbilgin}, another IM/DD based system FSO system which combines antenna shift keying with joint pulse position and amplitude modulations was considered.
For this system, which was denoted as spatial pulse position and amplitude modulation (SPPAM), the atmospheric turbulence channel was modeled as log-normal or Gamma-Gamma distributions and was evaluated, in terms of bounds,  for uncoded and coded signals. ABEP performance evaluation results have shown that SPPAM offers a compromise
between spectral and power efficiencies as well as a certain degree of robustness against atmospheric turbulence.
Despite these two papers which deal with non-coherent detection schemes, the potential enhancements of OSM on the performance of FSO systems with coherent detection
still remains an open research topic which, to the best of our knowledge, has not been addressed so far in the open technical literature.

Motivated by the above, in this paper we present for the first time a generic analytical framework which can be used to accurately obtain the performance of outdoor OSM with coherent detection in the presence of turbulence-induced fading.
More specifically and within this novel analytical framework, the main novel research contributions of the paper are as follows:
\begin{itemize}
  \item New analytical expressions for the ABEP of coherent OSM under turbulence conditions modeled by the H-K distribution are derived.
  When the transmitter is
equipped with two apertures the resulting analytical expressions are exact, whereas for an arbitrary number of transmit–
apertures tight upper–bounds
can be obtained.
    \item  Error probability performance bounds for coded OSM systems are derived and the performance enhancements when channel coding is employed are presented and analyzed.
\end{itemize}
The error probability performance of OSM is also compared to that of conventional FSO schemes with transmit or receive diversity only, i.e. when Maximal Ratio Combining (MRC),
Selection Combining (SC) or Alamouti-type Space-Time Block Codes (STBC) are employed.
It is noted that the theoretical analysis is substantiated by comparing the theoretical and equivalent simulated performance evaluation results obtained by means of Monte Carlo techniques.

The paper is organized as follows. After this introduction, Section~\ref{Sec:Model} outlines the system and channel models. In Section~\ref{Sec:ABEP_Analysis} analytical expressions for the ABEP of uncoded OSM systems are presented. Asymptotic ABEP expressions are also derived, wherefrom the diversity gain of coherent OSM can be readily deduced. The performance of coded OSM systems is discussed in Section~\ref{Sec:Coded}.
In Section~\ref{Sec:Results} the various performance evaluation results and their interpretations as well as comparisons are presented. Finally, concluding remarks can be found in Section~\ref{Sec:Conclusions}.
{\textit{Notations:} A comprehensive list of all mathematical notations used in this paper can be found in Table~\ref{Tab:Notations}.}

\section{System and Channel Model}\label{Sec:Model}
In this section, a detailed description of the OSM FSO system model , i.e. transmitter, channel and receiver is provided.
Moreover, the H-K distribution is introduced and analytical expressions
for its parameters in terms of equivalent physical parameters of the turbulence phenomenon, such as the refractive-index structure parameter,
optical wave number, and propagation path length, are derived.

\subsection{Preliminaries}
Let us consider a $M \times N$ MIMO FSO system with $M$ transmit units (lasers) and $N$ coherent receivers.
It is assumed that the receiving apertures are separated by more than a coherence wavelength to ensure the independency of fading channels.
The basic principle of OSM modulation is as follows \cite{J:OSM2011, J:FathHaas}:

i) The transmitter encodes blocks of $\log_2(M)$ data bits
into the index of a single transmit unit.
Such a block of bits is hereafter referred to as ``message"
and is denoted by $b_m$, $\forall m =1,2,...,M$.
It is assumed that the $M$ messages are transmitted with equal probability
by the encoder and that the related transmitted signal is
denoted by $\tilde{E}_{m} = E_{m}\exp(\jmath \phi_{b_m})$.
During each time slot, only one transmitter $\ell$, where $\ell = 1,2, \ldots , M$ is active for data
transmission. The information bits are
modulated on the electric field of an optical signal beam through an external modulator.
During this particular time slot, the remaining transmit lasers are kept silent, i.e. they do not transmit.

ii) At the receiver, the incoming optical field is mixed
with a local oscillator (LO) field and the combined wave
is first converted by the photodetector to an
electrical one. A bandpass filter is then employed to extract the intermediate frequency (IF) component
of the total output current. Finally, a $N$-hypothesis detection problem is solved to retrieve the active transmit unit index, which results in the estimation of the unique sequence of bits emitted by
the transmitter.

\subsection{Receiver Structure}
%The modulated electric field at the $m$-th transmitter output can be expressed as
%\begin{equation}
%e_m(t) = E_{0,m}\exp(\jmath \omega t + \phi_{t,m}),\,\, m = 1, \ldots, M
%\end{equation}
%where $E_{0,m}$ is the amplitude of the transmitted field, $\omega$ is the optical carrier frequency of the transmitters and
%$\phi_{t,m}$ is the phase noise from the $m$-th transmit laser.
The received electric field at the aperture plane of the $n$-th receiver
after mixing with a LO beam, can be expressed as \cite{J:Niu3, J:Bayaki2}
\begin{equation}
\begin{split}
e_n(t) & = \sqrt{2P_tZ_0}E_{m} h_{m,n} \cos(\omega_0 t + \phi_{m,n} + \phi_{b_m}) \\
& + \sqrt{2P_{LO}Z_0} \cos(\omega_{LO}t).
\end{split}
\end{equation}
In the above equation, $P_t$ is the transmit laser power, $Z_0$ is the free space impedance, $h_{m,n}$ and $\phi_{m,n}$
denote the magnitude and the phase of the complex channel coefficient between the $m$-th transmit and the $n$-th receive aperture, respectively.
Furthermore, $P_{LO}$ denotes the power of the local oscillator, $\omega_{LO} = \omega_0 + \omega_{IF}$ where $\omega_0$ and
$\omega_{IF}$ are the carrier and the intermediate radian frequencies, respectively.

The output current of the $n$-th photodetector can be mathematically expressed as \cite{J:Niu3, J:Bayaki2}
\begin{equation}\label{Eq:Photocurrent}
i_n(t) = \frac{R}{Z_0}[e_n(t)]^2
\end{equation}
where $R = \eta q_e /(h \nu_0)$ is the responsivity of the photodetector with $q_e = 1.6\times 10^{-19} $Cb is the charge of an electron, $h = 6.6\times 10^{-34} \rm{J\cdot s}$ is the Planck constant,
$\eta$ is the photodetector efficiency, and $\nu_0 = \omega_0/(2\pi)$ is the optical
center frequency.
Expanding \eqref{Eq:Photocurrent} and ignoring the double-frequency terms that are filtered out by the bandpass filter, the resulting photocurrent
can be expressed as
\begin{equation}\label{Eq:Photocurrent2}
\begin{split}
i_n(t) & = R P_t E_{m}^2 h_{m,n}^2+RP_{LO} \\
& +2R\sqrt{P_tP_{LO}}E_{m} h_{m,n}\cos(\omega_{IF} t - \phi_{m,n} - \phi_{b_m}) \\
& \triangleq i_{\rm{DC}}(t)+ i_{\rm{AC}}(t).
\end{split}
\end{equation}
In \eqref{Eq:Photocurrent2},
$i_{\rm{DC}}(t) \triangleq R P_t E_{m}^2 h_{m,n}^2+RP_{LO}$ is
the DC component generated by the signal and local oscillator fields, respectively, $i_{\rm{AC}}(t) \triangleq 2R\sqrt{P_tP_{LO}}\cos(\omega_{IF} t - \phi_{m,n} - \phi_{b_m})$
is the AC component in the received photocurrent which, unlike for direct detection, contains information about the frequency and phase of the
received signal. It is assumed that for coherent detection the intermediate frequency $\omega_{IF}$
is nonzero, so that the signal
power can be expressed as $P_s = 2R^2P_tP_{LO}E_{m}^2 h_{m,n}^2$

As in \cite{J:Bayaki2, J:Niu3, J:Aghajanzadeh, J:Kiasaleh}, we also consider that $P_{LO} \gg P_s$ and thus, the DC photocurrent can be approximated as
$i_{\rm{DC}}(t) \approx RP_{LO}$. The photodetection process is impaired by shot noise with variance
$\sigma_{\rm{shot,L}}^2 = 2q_eRP_{LO}B_e $ where $B_e$ is the electrical bandwidth of the photodetector.
It is also noted that because of the large value of $RP_{LO}$ the photocurrent due to thermal noise and the dark current can be ignored \cite{J:Niu3}.

%The average signal-to-noise ratio (SNR) of the $n$-th coherent receiving aperture in a given symbol interval can therefore be expressed as
%\begin{equation}\label{Eq:SNRcoherent}
%\gamma_{\rm{coherent} = \frac{\mathbb{E}\langle P_s \rangle}{\sigma_{\rm{shot,L}}^2} = \frac{RP_t|E_{m}|^2 \mathbb{E}\langle|h_{m,n}|^2\rangle}{eB_e}
%\end{equation}`
Following \cite{J:Bayaki2} and \cite{J:Aghajanzadeh}, the sufficient statistics at the $n$-th coherent receiver can be expressed as
\begin{equation}\label{Eq:signal_model}
y_n = \sqrt{\mu}h_{m,n}E_m\exp[\jmath(\phi_{m,n}+ \phi_{b_m})] + z_{n}
\end{equation}
where $\mu = {RP_t}/({q_eB_e})$ is the average signal-to-noise ratio (SNR) and $z_n$ is the noise at the $n$-th receiver. Assuming that the LO power is large and
the receiver noise is dominated by LO related noise
terms, the Additive White Gaussian Noise (AWGN) model can be employed as an accurate approximation of
the Poisson photon-counting detection model \cite{J:Bayaki2, J:Niu3}. Thus, $z_{n}$ can be modeled as a zero-mean unit variance complex Gaussian random variable
\cite{J:Bayaki2}.

Similar to \cite{J:RenzoRice}, it is assumed that
the receiver has knowledge of the actual fading
gains and that the total fading remains
constant over one bit interval and changes from one interval
to another in an independent manner.
At the receiver, the optimal spatial modulation detector estimates the active transmitter index, $\ell$, at a given time slot according to \cite{J:Jeganathan}
\begin{equation}\label{Eq:MLdetector}
\begin{split}
\hat{\ell}& = {\argmax_{\ell}} p_{\mathbf{y}}\left(\mathbf{y} | {\mathbf{x}}, {\mathbf{H}} \right) \\
& = {\argmin_{\ell}}\left\{\sqrt{\mu} \parallel \mathbf{h}_{\ell} x_{\ell}\parallel_F^2 -2\left(\mathbf{y}^{T}\mathbf{h}_{\ell}x_{\ell}\right)\right\}
\end{split}
\end{equation}
where
\begin{itemize}
  \item[-] $\mathbf{x}$ is an $M$-dimensional vector with elements corresponding to the electrical field ${E}_m\exp(\jmath\phi_{b_m})$ that is transmitted over the optical MIMO channel;
  \item[-] $\mathbf{H}(t)$ is an $N\times M$ optical MIMO channel defined as
\begin{equation}
\begin{split}
&\mathbf{H}(t) = [\mathbf{h}_{1}, \mathbf{h}_{2}, \ldots, \mathbf{h}_{M}] \\
& \triangleq \left[ \begin{array}{ccc}
{{h}}_{11}(t)\exp(\jmath\phi_{11})  & \ldots & {{h}}_{1M}(t)\exp(\jmath\phi_{1M})\\
{{h}}_{21}(t)\exp(\jmath\phi_{21})  & \ldots & {{h}}_{2M}(t)\exp(\jmath\phi_{2M})\\
\vdots  & \ddots & \vdots \\
{{h}}_{N1}(t)\exp(\jmath\phi_{N1})  & \ldots & {{h}}_{NM}(t)\exp(\jmath\phi_{NM})
\end{array} \right]\;
\end{split}
\end{equation}
\item[-] $\mathbf{z}$ is the $N$-dimensional noise vector;
\item[-] $p_{\mathbf{y}}\left(\mathbf{y} | {\mathbf{x}}, {\mathbf{H}} \right)$ is the PDF of $\mathbf{y}$ conditioned on the
transmitted vector $\mathbf{x}$ and the channel ${\mathbf{H}}$;
\end{itemize}

\subsection{Channel Model}
A discrete scattering model is considered, where the
radiation field of an optical wave at a particular point is assumed to be composed
of a number of scattered components that have traveled
different paths. Under the Ricean assumption \cite{J:AndrewsIK2}, the complex channel
path gains ${\tilde{h}}_{ij}(t)$  between the $i$-th transmitter and the $j$-th photodetector can be expressed as
${\tilde{h}}_{ij}(t) = h_{ij}(t)\exp(\jmath \omega t)$ where $\omega$ is the radian frequency of the optical wave
and
\begin{equation}\label{Eq:field}
\begin{split}
h_{ij}(t) & = \Re\{h_{ij}(t)\}+\jmath \Im\{h_{ij}(t)\}  \\
&= A_{ij}\exp[\jmath \theta_{ij}(t)]+R_{ij}(t)\exp[\jmath \Phi_{ij}(t)]
\end{split}
\end{equation}
where the term $A_{ij}\exp(\jmath \theta_{ij}(t))$ is a deterministic component and $R_{ij}(t)\exp(\jmath \Phi_{ij}(t))$
is a circular complex Gaussian random variable. Hence, the amplitude $R_{ij}$ is Rayleigh distributed with parameter $\sigma_{ij}^2 = b_{ij}/2$ \cite[Eq. (13)]{J:AndrewsIK2} and the phase $\Phi_{ij}$ is uniformly distributed over $[0, 2\pi)$.
Under the assumption of a doubly stochastic scintillation model \cite{J:AndrewsIK2},
the effect of random fluctuations in the turbulence parameters is
modeled by allowing random variations in the parameter $b_{ij}$ of the Rayleigh component.
Following \cite{J:AndrewsIK2}, it is further assumed that $b_{ij}$ follows a gamma distribution with PDF given by
\begin{equation}\label{Eq:pdfb}
f_{b_{ij}}(b) = \left(\frac{\alpha_{ij}}{b_0}\right)^{\alpha_{ij}}\frac{b^{\alpha_{ij}-1}}{\Gamma(\alpha_{ij})}\exp\left(-\frac{\alpha b}{b_{0_{ij}}}\right)
\end{equation}
where $\alpha$ is the shaping parameter and represent the effective number of scatters and $b_{0_{ij}} = \mathbb{E}\{b_{ij}\}$.
Then, the PDF of the irradiance $\mathrm{I}_{ij} = |{h}_{ij}(t)|^2$, $f_{\mathrm{I}_{ij}}(\mathrm{I})$, can be expressed as \cite[Eq. (8)]{J:AndrewsIK2}
\begin{equation}\label{Eq:PDFIK}
\begin{split}
& f_{\mathrm{I}_{ij}}(\mathrm{I}) = \frac{\left({\alpha_{ij}}/{b_{0_{ij}}}\right)^{\alpha_{ij}}}{\Gamma(\alpha_{ij})} \\
& \times \int_0^{\infty} b^{\alpha_{ij}-2}\exp\left(-\frac{\alpha_{ij} b}{b_{0_{ij}}}-\frac{\mathrm{I}+{A_{ij}}^2}{b}\right)I_0\left(\frac{2{A_{ij}}\sqrt{\mathrm{I}}}{b}\right)\mathrm{d}b
\end{split}
\end{equation}
which is actually the integral representation of the H-K distribution \cite{J:Jakeman2}.
It is noted that $f_{\mathrm{I}_{ij}}(\mathrm{I})$ cannot, in general, be expressed in closed form, with the exception of the special cases $A_{ij} = 0$ or $\alpha = 1$. Specifically, for $A_{ij} = 0$ \eqref{Eq:PDFIK} reduces to the K-distribution whereas for $\alpha = 1$, \eqref{Eq:PDFIK} reduces to a special case of the I-K distribution \cite[Eq. (10)]{J:Jakeman2}.

The $\nu$-th normalized moment of $\mathrm{I}_{ij}$ is given by \cite[Eq. (22)]{J:Jakeman2} as
\begin{equation}\label{Eq:moment}
\frac{\mathbb{E}\{\mathrm{I}_{ij}^\nu\}}{\mathbb{E}\{\mathrm{I}_{ij}\}^\nu} = \frac{\nu!}{\alpha_{ij}^\nu(1+\rho_{ij})^\nu}
\sum_{k=0}^{\nu}\binom{\nu}{k}\frac{\Gamma(\alpha_{ij}+\nu-k)}{\Gamma(\alpha_{ij})}\frac{(\alpha_{ij}\rho_{ij})^\nu}{\nu!}
\end{equation}
where $\rho_{ij} = A_{ij}^2/b_{0_{ij}}$ is the coherence parameter, defined as the power ratio of mean intensities of the constant-amplitude component and random component of
the field in \eqref{Eq:field} \cite{J:AndrewsIK2,J:AndrewsIK3}.
Using \eqref{Eq:moment}, the \emph{scintillation index} can be readily calculated as
\begin{equation}\label{Eq:SIHK}
\sigma_\mathrm{I_{ij}}^2 \triangleq \frac{\mathbb{E}\{\mathrm{I}_{ij}^2\}}{\mathbb{E}\{\mathrm{I}_{ij}\}^2}-1 =  \frac{\alpha_{ij}+2\alpha_{ij}\rho_{ij}+2}{\alpha_{ij}(1+\rho_{ij})^2}.
\end{equation}

Under the assumption of spherical wave propagation,
$\sigma_\mathrm{I_{ij}}^2$ can be directly related to atmospheric conditions as \cite[Eq. (7), Eq. (9)]{J:AndrewsIK3}
\begin{equation}\label{Eq:SI2}
\sigma_\mathrm{I_{ij}}^2 \approx
\begin{cases}
0.41\alpha_{ij}2(1 + 0.5\sigma_1^2),\, \sigma_1 \ll 1 \\
1+{2.8}/{\sigma_1^{{4}/{5}}},\, \sigma_1 \gg 1
\end{cases}
\end{equation}
where $\sigma_1^2=1.23 C_{n_{ij}}^2k^{7/6}L_{ij}^{11/6}$ is the Rytov variance, $k=2\pi/\lambda$ is the optical
wave number with $\lambda$ being the wavelength, $L_{ij}$ is the link distance and $C_{n_{ij}}$
denotes the index of refraction structure parameter. For FSO links near the
ground, $C_{n_{ij}}^2 \approx 1.7 \times 10^{-14} \mathrm{m}^{-2/3}$
and $8.4 \times 10^{-15} \mathrm{m}^{-2/3}$ for the daytime and night, respectively \cite{B:Goodman1985}.
Moreover, $\sigma_1 \ll 1$ and $\sigma_1 \gg 1$ correspond to weak and strong turbulence conditions, respectively.

Using \eqref{Eq:SI2}, the parameters of the H-K distribution, $\alpha$ and $\rho$, can be directly related to physical parameters of
the turbulence by following a similar line of arguments as in \cite{J:AndrewsIK3}, where similar results were derived for the I-K distribution. In particular, on the one hand, weak turbulence conditions are characterized in the
H-K distribution by large values of $\rho_{ij}$. In this case the scintillation index given by \eqref{Eq:SIHK}
can be approximated as
\begin{equation}\label{Eq:sigmaweak}
\sigma_\mathrm{I_{ij}}^2 \approx \frac{2}{\rho_{ij}}, \,\text{with } \rho_{ij} \gg 1.
\end{equation}
On the other hand, assuming strong turbulence conditions where $\rho_{ij}$ tends to zero, \eqref{Eq:SIHK} can be approximated as
\begin{equation}\label{Eq:sigmastrong}
\sigma_\mathrm{I_{ij}}^2 \approx 1+\frac{2}{\alpha_{ij}}, \,\text{with } \rho_{ij} \ll 1.
\end{equation}
By comparing \eqref{Eq:sigmaweak} and \eqref{Eq:sigmastrong} with the first and second branches of \eqref{Eq:SI2}, respectively,
$\alpha_{ij}$ and $\rho_{ij}$ can be obtained as
\begin{equation}\label{Eq:param-a}
\alpha_{ij} = 0.71\sigma_{1_{ij}}^{{4}/{5}}
\end{equation}
\begin{equation}\label{Eq:param-rho}
\rho_{ij} = \frac{4.88}{\sigma_{1_{ij}}^2(1+0.2\sigma_{1_{ij}}^2)}.
\end{equation}
To the best of our knowledge, the relationship of $\alpha_{ij}$ and $\rho_{ij}$ with $\sigma_{1_{ij}}$ given by \eqref{Eq:param-a} and \eqref{Eq:param-rho} is a novel result.
%Finally, it is pointed out that according to \cite{J:AndrewsIK2}, $f_{\mathrm{I}_{ij}}(\mathrm{I})$ can be accurately approximated with the PDF of the I-K distribution as
%\begin{equation}\label{Eq:CIK3}
%f_{\mathrm{I}_{ij}}(\mathrm{I}) \approx
%\begin{cases}
%2\alpha_{ij}(1+\rho_{ij})\left(\frac{(1+\rho_{ij})\mathrm{I}}{\rho_{ij}}\right)^{{\alpha_{ij}-1}/{2}}\\
%\times K_{\alpha_{ij}-1}(2\sqrt{\alpha_{ij}\rho_{ij}})\\
%         \times I_{\alpha_{ij}-1}\left(2\sqrt{\alpha_{ij}(1+\rho_{ij})\mathrm{I}}\right), \,\mathrm{I} < {\rho_{ij}}/({1+\rho_{ij}}) \\
%2\alpha_{ij}(1+\rho_{ij})\left(\frac{(1+\rho_{ij})\mathrm{I}}{\rho_{ij}}\right)^{{\alpha_{ij}-1}/{2}}\\
%\times I_{\alpha_{ij}-1}(2\sqrt{\alpha_{ij}\rho_{ij}})\\
%         \times K_{\alpha-1}\left(2\sqrt{\alpha_{ij}(1+\rho_{ij})\mathrm{I}}\right), \,\mathrm{I} < {\rho_{ij}}/({1+\rho_{ij}})
%\end{cases}
%\end{equation}
%Note that for the special case of $\alpha_{ij} = 1$, the approximation of \eqref{Eq:PDFIK} with \eqref{Eq:CIK3} becomes exact.

\section{Performance Analysis of Uncoded OSM}\label{Sec:ABEP_Analysis}
In this section, by employing the well-known MGF-based approach for the performance analysis of digital communications over fading channels \cite{B:Alouini}, analytical expressions for the ABEP of uncoded OSM systems will be derived. Expressions for the diversity and coding gains of OSM systems are also presented, thus providing useful insight as to how these parameters affect the overall system performance.
\subsection{Preliminaries}
For $M = 2$, the conditional bit error probability (BEP) of OSM systems when no turbulence induced fading is considered can be obtained in closed form as \cite{J:RenzoRice}
\begin{equation}\label{Eq:ABEP2xN}
P_E(\mathbf{h}_1, \mathbf{h}_2) = Q\left(\sqrt{\frac{\mu}{4}{\parallel \mathbf{h}_1- \mathbf{h}_2\parallel}_F^2 }\right).
\end{equation}
The squared Frobenius norm in \eqref{Eq:ABEP2xN} can be expressed as
\begin{equation}\label{Eq:Frob}
{\parallel \mathbf{h}_1- \mathbf{h}_2\parallel}_F^2  = \sum_{n=0}^N|h_{1,n}- h_{2,n}|^2
\end{equation}
where $h_{i,n}$ is the $n$-th element of $\mathbf{h}_i$, $\forall i \in \{1,2\}$.
When $M > 2$ transmitters are considered, a tight upper bound for the conditional BEP of the above system can be obtained as \cite[Eq. (7)]{J:OSM2011}
\begin{equation}\label{Eq:ABEPMxN}
\begin{split}
&P_E({\mathbf{H}}) \leq  \frac{M^{-1}}{\log_2(M)}\\
&\times  \sum_{m_1=1}^{M}\sum_{m_2 \neq m_1 =1}^{M}N_b(m_1, m_2){\rm PEP}(m_1\rightarrow m_2)
\end{split}
\end{equation}
where ${\rm PEP}(m_1\rightarrow m_2)$ denotes the pairwise error probability
(PEP) related to the pair of transmitters $m_1$ and $m_2$, where
$m_1$ and $m_2 \in 1, 2, \ldots, M$, and $N_b(m_1, m_2)$ is the number of bit which have occurred when the receiver decides incorrectly that $m_2$ instead of $m_1$ has been active.
%The PEP can be evaluated as \cite[Eq. (11)]{J:RenzoRice}
%\begin{equation}\label{Eq:ABEP2xNrPDF}
%{\rm PEP}(t_1\rightarrow t_2) = \int_0^{\infty}Q\left(\sqrt{\overline{\gamma}}z\right)f_{Z}(z)dz
%\end{equation}
The ${\rm PEP}(m_1\rightarrow m_2)$ can be evaluated as \cite[Eq. (8)]{J:OSM2011}
\begin{equation}\label{Eq:ABEP2xNrMGF}
{\rm PEP}(m_1\rightarrow m_2) = Q\left(\sqrt{\frac{\mu}{4}{\parallel \mathbf{h}_{m_1}- \mathbf{h}_{m_2}\parallel}_F^2 }\right).
\end{equation}
\subsection{MGF-Based Approach}
When atmospheric turbulence is taken into account, the conditional error probabilities in \eqref{Eq:ABEP2xN} and \eqref{Eq:ABEPMxN} need to be
averaged over the elements of the channel matrix $\mathbf{H}$ in order to evaluate the ABEP.
Without loss of generality, let us consider the case of a $2\times N$ MIMO system. Since $h_{i,n}$ are complex Gaussian random variables,
the difference $\Delta_n \triangleq h_{1,n}- h_{2,n}$ is a complex Gaussian random variable having mean equal to the difference of the means of $h_{i,n}$ and variance equal to the sum of variances of $h_{i,n}$.
In order to deduce a closed form expression for the ABEP, it is further assumed
that $h_{i,n}$ have uncorrelated real and imaginary components with the same variance
$\sigma_n^2 = b_n/2$. It is noted that such an assumption is justified for link distances of the order of km and for
aperture separation distances of the order of cm \cite{J:LeeChan, J:LetzepisHolland}. For example, in \cite{J:LetzepisHolland} it was reported that for a link
distance of 1.5 km, a wavelength of 1550 nm, an aperture
diameter of 1 mm and photodetectors separated by as little as 35 mm, which validates the independence assumption.

Consequently, $\Delta_n$ has uncorrelated components too and its squared envelope, $|\Delta_n|^2$,
is characterized by a non-central chi-square PDF as follows
\begin{align}\label{Eq:PDF1}
f_{|\Delta_n|^2 }(x|b_n) = \frac{1}{2b_n}\exp\left(-\frac{x+\tilde{A}_n^2}{2b_n}\right)I_0\left(\frac{\tilde{A}_n\sqrt{x}}{b_n}\right)
\end{align}
where $\tilde{A}_n = |A_{2,n}e^{\jmath\theta_{2,n}}-A_{1,n}e^{\jmath\theta_{1,n}}|$. Assuming that $b_n$ follows a gamma distribution with parameters $\alpha_n$ and $b_{0,n}$, the unconditional PDF of $|\Delta_n|^2$ is obtained by averaging \eqref{Eq:PDF1} with respect to $b_n$, i.e.
\begin{equation}\label{Eq:PDF2}
\begin{split}
& f_{|\Delta_n|^2}(x) = \frac{\left({\alpha_n}/{b_{0,n}}\right)^{\alpha_n}}{2\Gamma(\alpha_n)} \\
& \times \int_0^{\infty} b_n^{\alpha_n-2}\exp\left(-\frac{\alpha_n b_n}{b_{0,n}}-\frac{x+\tilde{A}_n^2}{2b_n}\right)I_0\left(\frac{\tilde{A}_n\sqrt{x}}{b_n}\right)\mathrm{d}b_n.
\end{split}
\end{equation}
As was pointed out in \cite{J:AndrewsIK2}, the integral in \eqref{Eq:PDF2} cannot be solved in closed form. Nevertheless, for the special case of
$\alpha_n = 1$, i.e. when one scatterer per branch is considered, and by employing \cite[Eq. (10)]{J:AndrewsIK2}, this integral can be evaluated in closed form as
\begin{equation}\label{Eq:CIK2}
f_{|\Delta_n|^2}(x) =
\begin{cases}
\frac{1}{b_{0,n}}K_0\left(\sqrt{{2\tilde{A}_n}/{b_{0,n}}}\right)I_0\left({\sqrt{2x}}/{b_{0,n}}\right),\,x < \tilde{A}_n^2 \\
\frac{1}{b_{0,n}}I_0\left(\sqrt{{2\tilde{A}_n}/{b_{0,n}}}\right)K_0\left({\sqrt{2x}}/{b_{0,n}}\right),\,x > \tilde{A}_n^2.
\end{cases}
\end{equation}
 Moreover, for the special case where $h_{1,n}$ and $h_{2,n}$ have identical mean value, i.e. when $\tilde{A}_n = 0$, \eqref{Eq:PDF2} yields the well known K-distribution with PDF given by
\begin{equation}\label{Eq:PDFKdistr}
\begin{split}
f_{|\Delta_n|^2}(x) & = {2^{{(1-\alpha_n)}/{2}}}{\Gamma(\alpha_n)}\left(\frac{\alpha_n x}{b_{0,n}}\right)^{{(\alpha_n-1)}/{2}} \\
& \times K_{\alpha_n-1}\left(\sqrt{\frac{2\alpha_n x}{b_{0,n}}}\right).
\end{split}
\end{equation}

By employing the MGF-based approach for the performance analysis of digital communications over fading channels, the average PEP (APEP) can be obtained as
\begin{equation}\label{Eq:APEPexact}
\mathrm{APEP} = \frac{1}{\pi}\int_{0}^{\pi/2}\prod_{n=1}^{N}\left[\mathcal{M}_{|\Delta_n|^2}\left(\frac{\mu}{8\sin^2\theta}\right)\right]\mathrm{d}\theta.
\end{equation}
Moreover, using the tight approximation for the Gaussian Q-function presented in \cite[Eq. (14)]{J:Chiani} (i.e., $Q(x) \approx {1}/{12}\exp(-x^2)+{1}/{4}\exp(-2x^2/3)$),
an expression accurately approximating APEP can be deduced as
\begin{equation}\label{Eq:APEPapprox}
\mathrm{APEP} \approx \frac{1}{12}\prod_{n=1}^{N}\left[\mathcal{M}_{|\Delta_n|^2}\left(\frac{\mu}{8}\right)\right]+\frac{1}{4}
\prod_{n=1}^{N}\left[\mathcal{M}_{|\Delta_n|^2}\left(\frac{\mu}{6}\right)\right].
\end{equation}
In the following analysis, analytical expressions for the MGF of $|\Delta_n|^2$ will be deduced. Specifically, the following result holds:
\newtheorem{proposition}{Proposition}
\begin{proposition}
An integral representation for the MGF of $|\Delta_n|^2$ can be deduced as
\begin{equation}\label{Eq:MGF}
\begin{split}
& \mathcal{M}_{|\Delta_n|^2}(s) = \frac{\left({\alpha_n}/{b_{0,n}}\right)^{\alpha_n}}{\Gamma(\alpha_n)} \\
& \times \int_0^{\infty}\frac{b^{\alpha_n-1}}{2bs+1}\exp\left(-\frac{\tilde{A}_ns}{2bs+1}-\frac{\alpha_n b}{b_{0,n}}\right)\mathrm{d}b.
\end{split}
\end{equation}
\end{proposition}
\begin{IEEEproof}
By employing the definition of the MGF, $\mathcal{M}_{|\Delta_n|^2}(s)$ can be obtained as
\begin{equation}
\begin{split}
& \mathcal{M}_{|\Delta_n|^2}(s) = \int_0^{\infty}\exp(-sx)f_{|\Delta_n|^2}(x)\mathrm{d}x \\
& = \frac{\left({\alpha_n}/{b_{0,n}}\right)^{\alpha_n}}{2\Gamma(\alpha_n)}\int_0^{\infty}\int_0^{\infty} \exp\left(-sx-\frac{\alpha_n b}{b_{0,n}}-\frac{x+\tilde{A}_n^2}{2b}\right) \\
& \times I_0\left(\frac{\tilde{A}_n\sqrt{x}}{b}\right)b^{\alpha_n-2}\mathrm{d}b\mathrm{d}x.
\end{split}
\end{equation}
By changing the order of integration, the above equation can be expressed as
\begin{equation}\label{Eq:MGFinterm1}
\begin{split}
& \mathcal{M}_{|\Delta_n|^2}(s) = \frac{\left({\alpha_n}/{b_{0,n}}\right)^{\alpha_n}}{2\Gamma(\alpha_n)}\int_0^{\infty}b^{\alpha_n-2}
\exp\left(-\frac{\alpha_n b}{b_{0,n}}\right)\\
&\left[\int_0^{\infty} \exp\left(-sx-\frac{x+\tilde{A}_n^2}{2b}\right)I_0\left(\frac{\tilde{A}_n\sqrt{x}}{b}\right)\mathrm{d}x\right] \mathrm{d}b.
\end{split}
\end{equation}
The inner integral, i.e. with respect to $x$ can be evaluated by employing \cite[Eq. (3.15.2.2)]{B:Prudnikov4} as
\begin{equation}\label{Eq:MGFinterm2}
\begin{split}
& \int_0^{\infty} \exp\left(-sx-\frac{x+\tilde{A}_n^2}{2b}\right)I_0\left(\frac{\tilde{A}_n\sqrt{x}}{b}\right)\mathrm{d}x = \\
& \frac{2b}{2sb+1}\exp\left[\frac{1}{2\tilde{A}_nb(2sb+1)}\right].
\end{split}
\end{equation}
Substituting \eqref{Eq:MGFinterm2} into \eqref{Eq:MGFinterm1} and after some straightforward manipulations, \eqref{Eq:MGF} is readily deduced thus completing the mathematical proof.
\end{IEEEproof}
The integral in \eqref{Eq:MGF} can be accurately approximated by employing a Gauss-Chebyshev Quadrature (GCQ) technique as \cite{B:Abramowitz}
\begin{equation}
\begin{split}
& \mathcal{M}_{|\Delta_n|^2}(s) \approx \frac{\left({\alpha_n}/{b_{0,n}}\right)^{\alpha_n}}{\Gamma(\alpha_n)} \\
& \times \sum_{j=0}^Jw_j\frac{{t_j}^{\alpha_n-1}}{2{t_j}s+1}\exp\left(-\frac{\tilde{A}_ns}{2{t_j}s+1}-\frac{\alpha_n {t_j}}{b_{0,n}}\right)
\end{split}
\end{equation}
where $J$ is the number of integration points, $t_j$ are the abscissas and $w_j$ the corresponding weights. In \cite[eqs. (22) and (23)]{C:YilmazMGF}, $t_j$ and $w_j$ are defined as
\begin{subequations}\label{Eq:GCQ}
\begin{equation}\label{Eq:xk}
t_j = \tan\left[\frac{\pi}{4}\cos\left(\frac{2j-1}{2J}\pi\right)+\frac{\pi}{4}\right]
\end{equation}
\begin{equation}\label{Eq:wk}
w_j = \frac{\pi^2\sin\left(\frac{2j-1}{2J}\pi\right)}{4J\cos^2\left[\frac{\pi}{4}\cos\left(\frac{2j-1}{2J}\pi\right)+\frac{\pi}{4}\right]}.
\end{equation}
\end{subequations}
For the special case of $\tilde{A}_n = 0$, it can be shown that \eqref{Eq:MGF} can be evaluated in closed form. Specifically, the following result holds:
\newtheorem{corrolary}{Corrolary}
\begin{corrolary}
For the special case of $\tilde{A} = 0$ the MGF of $|\Delta_n|^2$ can be deduced in closed form as
\begin{equation}\label{Eq:MGFK}
\begin{split}
\mathcal{M}_{|\Delta_n|^2}(s) & = \left(\frac{\alpha_n}{2sb_{0,n}}\right)^{\frac{\alpha_n}{2}}\exp\left(\frac{\alpha_n}{4sb_{0,n}}\right)\\
& \times W_{-\frac{\alpha_n}{2}, \frac{\alpha_n-1}{2}}\left(\frac{\alpha_n}{2sb_{0,n}}\right).
\end{split}
\end{equation}
\end{corrolary}
This result can be readily deduced by employing the integral representation of the Whittaker $W$-function given in \cite[Eq. (9.222)]{B:Gradshteyn_00}. Moreover it is worth pointing out that \eqref{Eq:MGFK} is in agreement with a previously known result, namely the analytical expression for the MGF of the K-distribution.
\cite[Eq. (4)]{J:Theofilakos_GSC}.

\subsection{Analysis of the Diversity Gain}
The diversity gain of the considered OSM MIMO
system can be obtained by using the approach presented
in \cite{J:WangGiannakis}.
In particular, a generic analytical expression, which becomes asymptotically tight at high SNR values,
will be derived for the ${\rm APEP}$ appearing in \eqref{Eq:APEPexact}, as follows:
\begin{proposition}
For high SNR values, \eqref{Eq:APEPexact} can be approximated by
\begin{equation}\label{Eq:PEPasym}
{\rm APEP} \overset{\mu \gg 1}{\approx} \frac{2^{N - 1}\Gamma\left(N + \frac{1}{2}\right)}{\sqrt{\pi}\Gamma\left(N + 1\right)} \left[\prod_{n=1}^{N} c_\ell\right] \left(\frac{\mu}{4}\right)^{-N}
\end{equation}
where
\begin{equation}\label{Eq:codinggain}
c_n = \left(\frac{\tilde{A}_n}{2}\right)^{\frac{\alpha_n-1}{2}} \frac{\left(\alpha_n/b_{0,n}\right)^{\frac{\alpha_n+1}{2}}}{\Gamma(\alpha_n)}K_{\alpha_n-1}\left(\sqrt{\frac{2\tilde{A}_n\alpha_n}{b_{0,n}}}\right).
\end{equation}
\end{proposition}
\begin{IEEEproof}
According to \cite[Proposition 3]{J:WangGiannakis}, the asymptotic error performance of the OSM system
depends on the behavior of $\mathcal{M}_{|\Delta_n|^2}(s)$, as $s\rightarrow\infty$.
To determine an analytical asymptotic expression for APEP a Taylor series expansion is employed to approximate $\mathcal{M}_{|\Delta_n|^2}(s)$ as
\begin{equation}\label{Eq:MGFTaylor}
|\mathcal{M}_{|\Delta_n|^2}(s)| = c_n|s|^{-d_n}+o(|s|^{-d_n}), \, s\rightarrow\infty
\end{equation}
where $c_n$ and $d_n$ are parameters that determine the diversity and coding gains of the $n$-th diversity branch, respectively.
Observe that since ${\tilde{A}s}/({2sb+1})\overset{s\rightarrow\infty}{\approx}{\tilde{A}}/({2b})$ and
${1}/({2sb+1})\overset{s\rightarrow\infty}{\approx}1/({2bs})$,
\eqref{Eq:MGF} yields
\begin{equation}\label{Eq:MGF2}
\begin{split}
 \mathcal{M}_{|\Delta_n|^2}(s) & \approx \frac{\left({\alpha_n}/{b_{0,n}}\right)^{\alpha_n}}{2s\Gamma(\alpha_n)} \\
& \times  \int_0^{\infty}b^{\alpha_n-2}\exp\left(-\frac{\tilde{A}_n}{2b}-\frac{\alpha_n b}{b_{0,n}}\right)\mathrm{d}b.
\end{split}
\end{equation}
By employing \cite[Eq. (2.2.2.1)]{B:Prudnikov4}, \eqref{Eq:MGF2} can be solved in closed form yielding
\begin{equation}\label{Eq:MGF3}
\begin{split}
\mathcal{M}_{|\Delta_n|^2}(s) & \approx \left(\frac{\tilde{A}_n}{2}\right)^{\frac{\alpha_n-1}{2}} \frac{\left(\alpha_n/b_{0,n}\right)^{\frac{\alpha_n+1}{2}}}{s\Gamma(\alpha_n)} \\
& \times K_{\alpha_n-1}\left(\sqrt{\frac{2\tilde{A}_n\alpha_n}{b_{0,n}}}\right).
\end{split}
\end{equation}
By comparing \eqref{Eq:MGF3} and \eqref{Eq:MGFTaylor} it is readily deduced that $d_n = 1$ and $c_n$ is given by \eqref{Eq:codinggain}.
Thus, by substituting \eqref{Eq:MGFTaylor} into \eqref{Eq:APEPexact}, the asymptotic PEP expression can be obtained as
in \eqref{Eq:PEPasym} which concludes the proof.
\end{IEEEproof}
From \eqref{Eq:PEPasym} it is clear that the diversity gain achieved by the considered system is equal to $N$.
It is also evident that the diversity gain depends only on the number of the receive apertures and is independent of the fading severity.
This finding is in agreement with relevant findings reported in \cite{J:RenzoRice} and \cite{J:RenzoAsym}, for the case of radio-frequency MIMO wireless systems.

It is noted that for the special case $\tilde{A}_n = 0$, i.e. when $|\Delta_n|^2$ follows the K-distribution, by employing the asymptotic result $K_t(x) \overset{x\rightarrow 0}{\approx} ({\Gamma(t)}/{2})\left({2}/{x}\right)^t$ \cite{B:Abramowitz},
$c_n$ can be further simplified as
\begin{equation}
c_n = \frac{\alpha_n}{2b_{0,n}(\alpha_n-1)}.
\end{equation}
%\newcounter{mytempeqncnt}
%%_{e_\ell [r]}%
%\begin{figure*}[!t]
%\normalsize \setcounter{mytempeqncnt}{\value{equation}}
%\setcounter{equation}{10}
%\begin{equation}\label{Eq:MGFIK}
%\begin{split}
%\mathcal{M}_{I_n^2}(s) & = \left(\frac{1+\rho_n}{\rho_n}\right)^{(\alpha_n-1)/{2}}
%K_{\alpha_n-1}\left(2\sqrt{\alpha_n\rho_n}\right)\sum_{j=0}^{\infty}s^{-(\alpha_n+j)/{2}}\gamma\left(\frac{\alpha_n+j}{2}, A_n^2s\right)
%\frac{\left[\alpha_n(1+\rho_n)\right]^{j+(\alpha_n+1)/{2}}}{j!\Gamma(\alpha_n+j)} \\
%& + \frac{\pi}{2\sin(\alpha_n-1)\pi}\left(\frac{1+\rho_n}{\rho_n}\right)^{(\alpha_n-1)/{2}}
%K_{\alpha_n-1}\left(2\sqrt{\alpha_n\rho_n}\right)\sum_{j=0}^{\infty}\left\{
%\frac{\left[\alpha_n(1+\rho_n)\right]^{j+(3-\alpha_n)/{2}}}{j!\Gamma(j-\alpha_n+2)}
%s^{-\frac{1+j}{2}}\Gamma\left(\frac{1+j}{2}, A_n^2s\right)\right. \\
%& \left. \frac{\left[\alpha_n(1+\rho_n)\right]^{j+(1+\alpha_n)/{2}}}{j!\Gamma(j+\alpha_n)}
%s^{-(\alpha_n+j)/{2}}\Gamma\left(\frac{\alpha_n+j}{2}, A_n^2s\right) \right\}
%\end{split}
%\end{equation}
%\hrulefill \vspace*{1pt} \setcounter{equation}{\value{mytempeqncnt}}
%\end{figure*}

\section{Performance Analysis of Coded OSM over turbulence channels}\label{Sec:Coded}
%In this section, new upper bounds for the ABEP of coded OSM systems are derived, using the transfer function method.
When coded OSM is employed, the input signal $\mathbf{s}(t)$ is first encoded by a convolutional
encoder. The encoded data are interleaved by a random block
interleaver and transmitted through the optical wireless channels using spatial modulation. It is also assumed that perfect
interleaving at the transmitter and de-interleaving at the receiver
is used.
Assuming maximum likelihood soft decision decoding,
the log likelihood ratios (LLRs) for the $i$-th constellation bit when the $\ell$-th transmitting antenna is active
are computed as \cite[Eq. (6)]{J:OSM2011}
\begin{equation}
\begin{split}
\mathrm{LLR} & = \log\frac{\mathrm{Pr}\{\ell^i= 1| \mathbf{y}\}}{\mathrm{Pr}\{\ell^i= 0| \mathbf{y}\}} \\
& = \log\frac{\sum_{\hat{\ell} \in \mathcal{L}_1^i}\exp\left(-{\parallel \mathbf{y}-\mathbf{h}_{\hat{\ell}}s_\ell \parallel^2}/{N_0}\right)}{\sum_{\hat{\ell} \in \mathcal{L}_0^i}\exp\left(-{\parallel \mathbf{y}-\mathbf{h}_{\hat{\ell}}s_\ell \parallel^2}/{N_0}\right)}
 \end{split}
\end{equation}
where $\mathcal{L} \in \{1 : M\}$ is the set of spatial constellation points,
$\mathcal{L}_1^i$ and $\mathcal{L}_0^i$ are subsets from $\mathcal{L}$ containing the transmitter
indices having "1" and "0" at the $i$-th bit, respectively.
The resulting data are finally decoded by a Viterbi decoder.

A union bound on the ABEP of a coded communication system can be evaluated as \cite{B:Alouini}
\begin{equation}\label{Eq:coded1}
\bar{P}_{\mathrm{ub}} \leq \frac{1}{n}\sum_{\mathbf{X}}P(\mathbf{X})\sum_{\mathbf{X}\neq \mathbf{X'}}q(\mathbf{X},\mathbf{X'})\mathrm{PEP}(\mathbf{X},\mathbf{X'})
\end{equation}
where $P(\mathbf{X})$ is the probability that the coded sequence $\mathbf{X}$ is
transmitted, $q(\mathbf{X},\mathbf{X'})$ is the number of information bit errors
in choosing another coded sequence $\mathbf{X'}$ instead of $\mathbf{X}$ $n$ is
the number of information bits per transmission and $\mathrm{PEP}(\mathbf{X},\mathbf{X'})$ is the pairwise error probability, i.e the probability of selecting $\mathbf{X'}$
when $\mathbf{X}$ was actually transmitted.

By employing \cite[p. 510]{B:Alouini}, \eqref{Eq:coded1} can be efficiently evaluated as
\begin{equation}\label{Eq:bound1}
\bar{P}_{\mathrm{ub}} \leq \frac{1}{n}\sum_{\mathbf{X}}P(\mathbf{X})\int_0^{\pi/2}\left[\left.\frac{\partial }{\partial N}T[D(\theta), N]\right|_{N=1}\right]
\end{equation}
where $T[D(\theta), N]$ is the transfer function of the employed convolutional code, $N$ is an indicator variable taking into account the number
of the erroneous bits and $D(\theta)$ depends on the underlying PEP expression.
Furthermore, assuming that uniform error probability (UEP) codes are considered and taking into account the symmetry property this code family exhibits,
thus making the distance structure of a UEP code independent of the transmitted sequence, \eqref{Eq:bound1} can be further simplified as \cite{B:Alouini}
\begin{equation}\label{Eq:bound2}
\bar{P}_{\mathrm{ub}} \leq \frac{1}{\pi}\int_0^{\pi/2}\left[\frac{1}{n}\left.\frac{\partial }{\partial N}T[D(\theta), N]\right|_{N=1}\right].
\end{equation}
For $M=2$, using \eqref{Eq:ABEP2xN}, \eqref{Eq:Frob} and Craig's formula for the Gauss Q-function, i.e. $Q(x) = 1/\pi\int_0^{\pi/2}\exp(-x^2/2\sin^2\theta)\mathrm{d}\theta$, $D(\theta)$
can be expressed as
\begin{equation}
D(\theta) = \prod_{n=1}^N\mathcal{M}_{|\Delta_n|^2}\left(\frac{\mu}{8\sin^2\theta}\right)
\end{equation}
where $\mathcal{M}_{|\Delta_n|^2}$ can be obtained from \eqref{Eq:MGF}.
When $M>2$, by employing \cite[Eq. (13)]{J:OSM2011},
and using a similar line of arguments as in the case of $M = 2$, $D(\theta)$ can be written as
\begin{equation}
\prod_{m_1=1}^{M}\prod_{m_2 \neq m_1 =1}^{M}\mathcal{M}_{|\Delta_{m_1, m_2}|^2}\left(\frac{\mu}{8\sin^2\theta}\right)
\end{equation}
where $|\Delta_{m_1, m_2}|^2 = \parallel \mathbf{h}_{m_1} -\mathbf{h}_{m_2} \parallel^2$. The last MGF can be easily computed analytically with the help of \eqref{Eq:MGF}.

\section{Performance Evaluation Results and Discussion}\label{Sec:Results}
\begin{figure}[!t]
\centering
\includegraphics[keepaspectratio,width=3.0in]{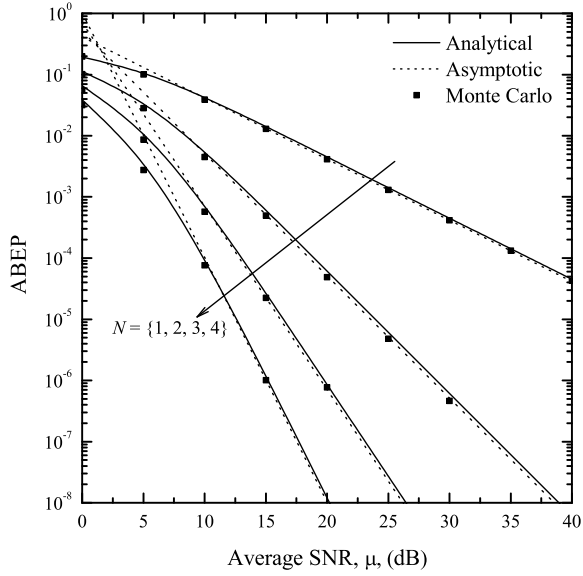}
\caption{ABEP of uncoded OSM for $2\times N$  MIMO H-K turbulent channels as a function of the average SNR, $\mu$, for various number of receiving apertures, $N$.
Simulation Parameters: $A_{1,n} = 2$, $A_{2,n} = 1$, $\theta_{1,n} = \pi/3$, $\theta_{2,n} = \pi/4$, $\alpha_{n} = 2$, $b_{0,n} = 2$.
} \label{Fig:ABEP1}
\end{figure}
\begin{figure}[!t]
\centering
\includegraphics[keepaspectratio,width=3.0in]{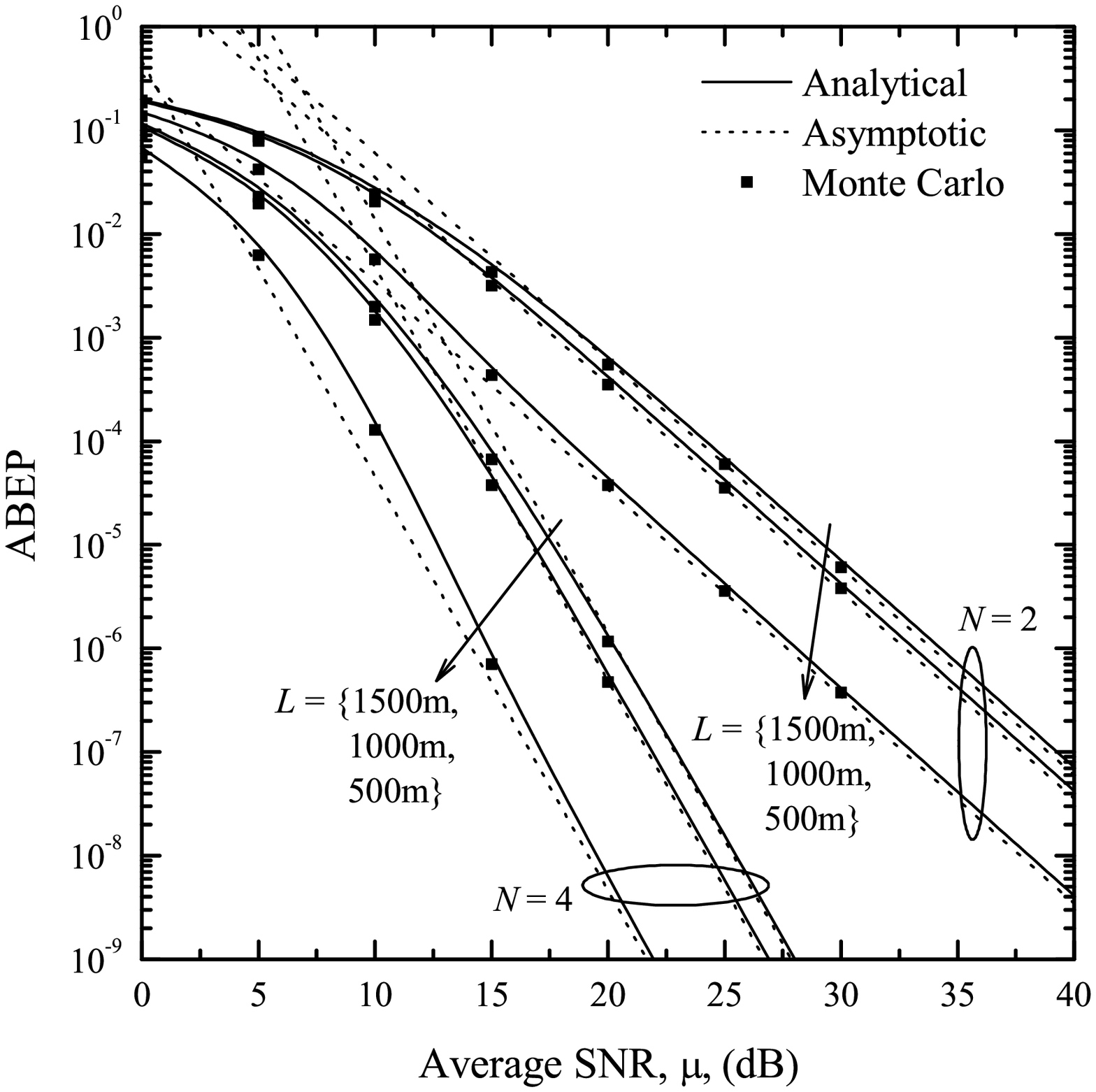}
\caption{ABEP of uncoded OSM for $2\times2$ and $2\times4$ MIMO H-K turbulent channels as a function of the average SNR, $\mu$, for various values of link distances, $L$.
Simulation Parameters: $\lambda = 1550{\rm nm}$, $C_n^2 = 1.7\times 10^{-14} {\rm m}^{-2/3}, \theta_{1,n} = \pi/3, \theta_{2,n} = \pi/4$.
} \label{Fig:ABEP2}
\end{figure}
\begin{figure}[!t]
\centering
\includegraphics[keepaspectratio,width=3.0in]{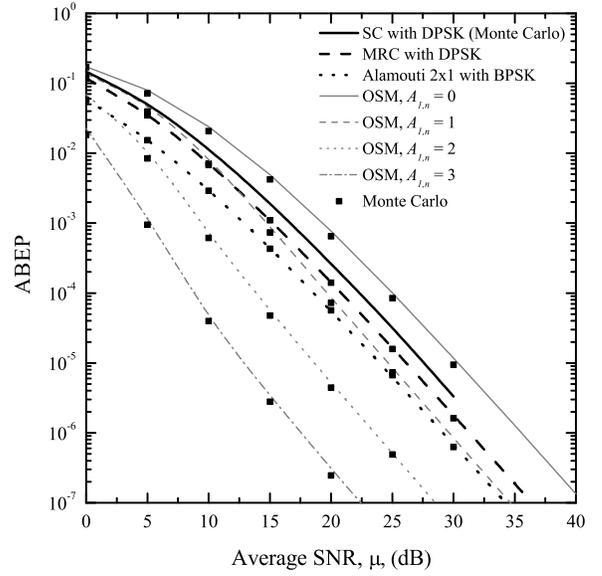}
\caption{ABEP Comparison of $2\times2$ OSM with $1\times2$ coherent MRC systems employing DPSK, as a function of the average SNR, $\mu$, for various values of $A_{1,n}$.
Simulation Parameters: $A_{2,n} = 0$, $\theta_{1,n} = 0$, $\theta_{2,n} = 0$, $\alpha_{n} = 1.5$, $b_{0,n} = 1.5$.
} \label{Fig:MRCComparison}
\end{figure}
\begin{figure}[!t]
\centering
\includegraphics[keepaspectratio,width=3.0in]{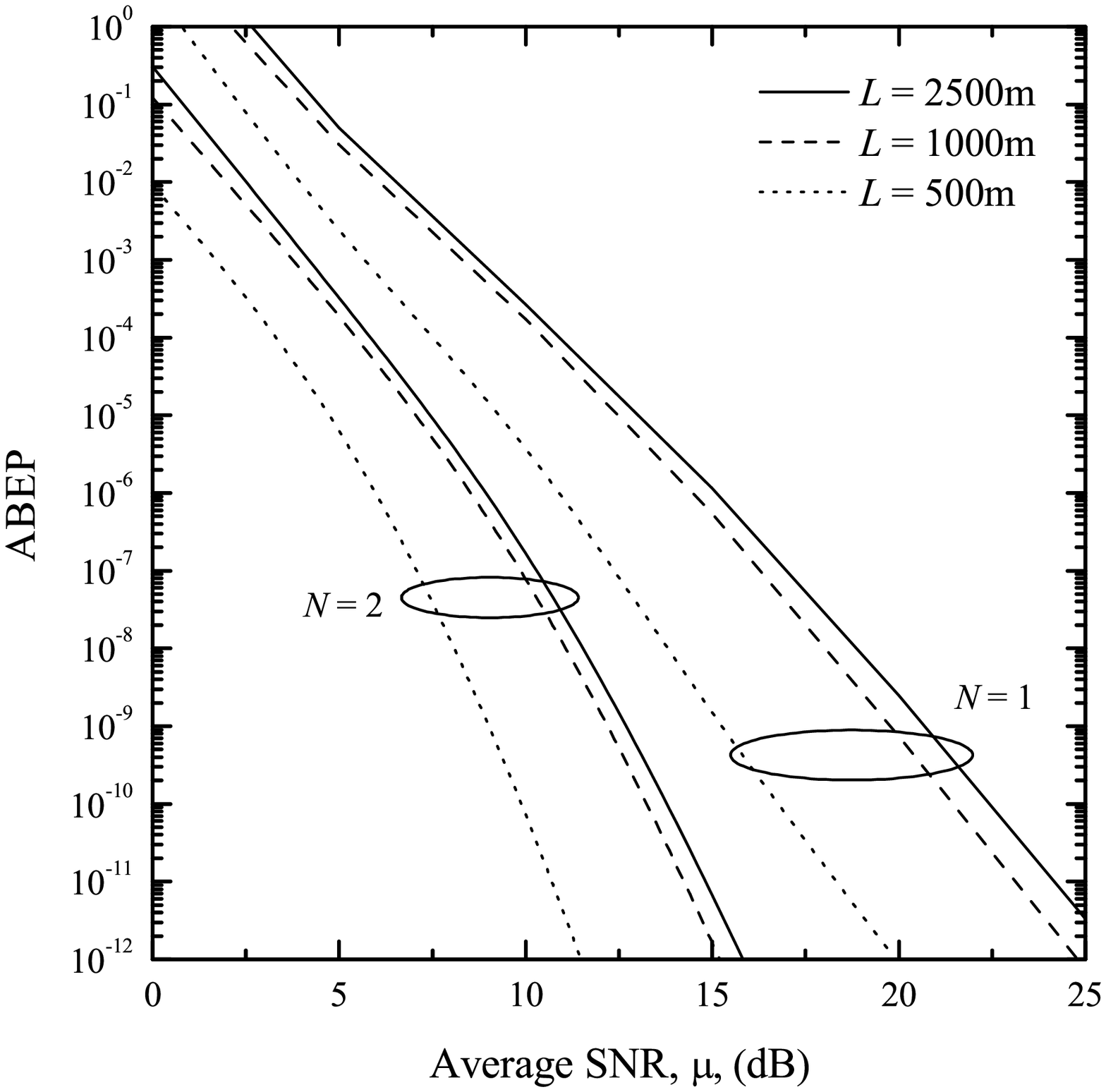}
\caption{ABEP upper bounds of convolutional coded OSM for $2\times2$ and $2\times1$ H-K turbulent channels as a function of the average SNR, $\mu$, for various values of link distances, $L$.
Simulation Parameters: $\lambda = 1550{\rm nm}$, $C_n^2 = 1.7\times 10^{-14} {\rm m}^{-2/3}$, $\theta_{1,n} = \pi/3$, $\theta_{2,n} = \pi/4$.
} \label{Fig:Coded}
\end{figure}
In this section the various performance evaluation results which have been obtained by numerically evaluating the mathematical expressions presented in Sections  \ref{Sec:ABEP_Analysis} and \ref{Sec:Coded} for uncoded and coded OSM systems operating over H-K turbulent channels will be presented.
In particular, for uncoded OSM systems the following performance evaluation results have been obtained: \textit{i}) ABEP vs. SNR for $2\times N_r$ OSM systems (obtained using \eqref{Eq:APEPapprox} with \eqref{Eq:MGF}, and \eqref{Eq:PEPasym} - see Figs.~\ref{Fig:ABEP1}, \ref{Fig:ABEP2} and \ref{Fig:MRCComparison});
\textit{ii}) ABEP vs. SNR for $2\times N$ MIMO OSM systems, $2\times N$ MIMO
(obtained using \eqref{Eq:APEPapprox} with \eqref{Eq:MGF}).
For the uncoded schemes, in order to validate the accuracy of the previously mentioned expressions, comparisons with complementary Monte Carlo simulated performance results are also included in these figures.
As far as the performance of coded OSM systems is concerned, ABEP upper bounds vs. SNR have been obtained using \eqref{Eq:bound2} with \eqref{Eq:MGF} (see Fig.~\ref{Fig:Coded}).

Fig.~\ref{Fig:ABEP1}, presents the ABEP performance as a function of the average SNR, $\mu$, of $2\times N$ MIMO OSM systems with
$N \in \{1,2,3,4\}$. Independent and identically distributed branches are considered with
$A_{1,n} = 2$, $A_{2,n} = 1$, $\theta_{1,n} = \pi/3$, $\theta_{2,n} = \pi/4$, $\alpha_{n} = 2$, $b_{0,n} = 2$.
The obtained results clearly indicate that the ABEP curves, obtained using \eqref{Eq:APEPapprox}, are in close agreement with those obtained via simulations, verifying the correctness of the proposed analysis. Moreover, it is evident that the asymptotic ABEP curves correctly predict the diversity gain of the considered system for all tested cases.

In Fig.~\ref{Fig:ABEP2}, the dependence on the link distance $L$ of the ABEP of a $2\times N$ MIMO OSM system
is illustrated. The considered system is again equipped with either $N = 2$ or $N = 4$ receiving apertures and identically distributed branches are assumed.
The parameters of the H-K distribution are calculated from \eqref{Eq:param-a} and \eqref{Eq:param-rho}
assuming spherical wave propagation. Following \cite{J:Uysal}, it is further assumed that
the operating wavelength is $\lambda = 1550$ nm and $C_{n}^2 = 1.7 \times 10^{-14} \mathrm{m}^{-2/3}$.
As expected, the error performance deteriorates as $L$ increases from $L=500$m to $L=1500$m.
Moreover, it is evident that an increase in $L$ from 500m to 1000m results in a more severe performance deterioration than in the case where
$L$ increases from 1000m to 1500m. In all cases considered, the analytical results obtained using \eqref{Eq:APEPapprox} are compared with the equivalent results obtained by means of Monte-Carlo computer simulations and again match very well.

Next we compare the proposed OSM system with two alternative coherent FSO systems that can provide performance enhancements by means of transmit
(MISO) or receive diversity (SIMO). It is noted that for similar aperture configurations,
a fair comparison between coherent and IM/DD systems seems difficult as the same received laser power leads to different SNRs for each of these
schemes \cite{J:Bayaki2}. On the other hand, in order to perform a fair comparison between OSM and the alternative MISO or SIMO systems under the same propagation channel conditions,
the aperture configuration of the FSO systems under comparison should be selected
carefully. Specifically, because of the fact that the diversity gain of OSM equals
to only the number of the receive apertures only, i.e. no transmit diversity gain is
provided, the number of transmit or receive apertures of the alternative systems must
be hence selected to be equal to the number of receive apertures of the OSM system.
To this end, for a fair comparison in our paper a $2\times2$ OSM system is compared with
 the following two alternative FSO communication systems which also employ coherent detection:
i) A $1\times2$ heterodyne FSO communication system which employs Differential
Phase Shift Keying \cite{J:Kiasaleh} and MRC or SC.

ii) A $2\times1$ coherent FSO system employing the Alamouti scheme \cite{J:Niu3} and Binary
Phase Shift Keying (BPSK).

%In the following, OSM system with two other alternative (well established) coherent FSO systems that can provide performance enhancements by means of transmit or receive diversity. It is noted that for similar aperture configurations, a fair comparison between coherent and IM/DD systems is, in general, difficult to be performed as the same received laser power leads to different SNRs for both schemes \cite{J:Bayaki2}. On the other hand, in order to perform a fair comparison between OSM and the alternative considered systems under the same propagation conditions, the aperture configuration of all considered systems should be selected in a suitable manner. Specifically, because of the fact that the diversity gain of OSM is equal to the number of the receive apertures only (i.e. no transmit diversity gain is provided), the number of transmit or receive apertures of the alternative systems is hereafter selected to be equal to the number of receive apertures of the OSM system.
%To this end, a $2\times 2$ OSM system is compared with
%i)	A $1\times 2$ heterodyne FSO communication system which employs Differential Phase Shift Keying (DPSK) \cite{J:Kiasaleh} and Maximal Ratio Combining (MRC) or Selection Combining (SC).
%ii)	A $2\times 1$ coherent FSO system employing the Alamouti scheme and Binary Phase Shift Keying (BPSK) \cite{J:Niu3}.

The instantaneous SNR at the output of the coherent MRC receiver assuming equal average SNR per receiving aperture, $\mu$ can be expressed as \cite{B:Alouini}
\begin{equation}\label{Eq:MRC}
\gamma_{\rm{MRC}} = \mu\sum_{n=1}^NI_n
\end{equation}
whereas for SC is
\begin{equation}
\gamma_{\rm{SC}} = \max\{ \mu I_1, \mu I_2\}.
\end{equation}
%The conditioned on the fading intensity BEP of the coherent DPSK FSO system can be expressed as \cite{J:Kiasaleh}
%\begin{equation}
%P(e|I) =\frac{1}{2}\exp(-\gamma)
%\end{equation}
%where $\gamma$
For MRC case, the ABEP can be deduced as \cite{B:Alouini}
\begin{equation}
P_E =\frac{1}{2}\prod_{n=1}^N\mathcal{M}_{I_n}(\mu).
\end{equation}
For SC case, an analytical expression for the ABEP is more difficult to be deduced and, therefore, ABEP will be evaluated by means of Monte Carlo simulation only.

As far as the Alamouti scheme is concerned, the instantaneous SNR at the input of the demodulator of the optical receiver has a similar form as \eqref{Eq:MRC} \cite{J:Niu3}.
For this scheme, the ABEP of BPSK can be evaluated as
\begin{equation}
P_E =\frac{1}{\pi}\int_0^{\pi/2}\prod_{n=1}^N\mathcal{M}_{I_n}\left(\frac{\mu}{\sin^2\theta}\right)\mathrm{d}\theta.
\end{equation}
In order to simplify the underlying mathematical analysis, it is assumed that the PDF of ${I_n}$ is given by \eqref{Eq:PDFIK} with the parameters $A_n$ being all zero, i.e. the PDF is the K-distribution.
Thus, $\mathcal{M}_{I_n}(\mu)$ can be readily obtained in closed form from \eqref{Eq:MGFK} by replacing $b_{0,n}$ with $b_{0,n}/2$.
In Fig.~\ref{Fig:MRCComparison}, the ABEP of $2\times 2$ MIMO OSM links
is compared with the ABEP of $1 \times 2$ coherent FSO systems with DPSK.
and identically distributed links are considered.
In order to compare these systems under the same propagation conditions, it is assumed that $\alpha_n =1,5$, $b_{0,n} = 1.5$, $A_{2,n} = 0$ and $A_{1,n} = \{0,1,2,3\}$. As it can be observed, when either MRC or SC are employed, although coherent DPSK performs worse than OSM for values of $A_{1,n}$ up to approximately 1, it outperforms OSM at lower values of $A_{1,n}$. Moreover, although the OSM outperforms the Alamouti scheme for $A_{1,n} = 2$ and 3, it performs similarly for high SNR values when $A_{1,n} = 1$.
It is noted that for $A_{1,n} = 1$ and lower values of $A_{1,n}$ the Alamouti scheme yields the best performance of the considered OSM schemes.
When more transmit appertures are employed, however, this advantage is compensated by the superior spectral efficiency of OSM and its lower hardware complexity as compared to coherent MRC. Specifically, as pointed out in \cite{J:OSM2011}, OSM offers increased spectral efficiency by a factor $\log_2(M)$. Moreover, as only one transmitting aperture
is activated at any symbol duration, OSM has a lower decoding complexity as compared to conventional MRC and Alamouti schemes.

In Fig.~\ref{Fig:Coded}, upper bounds on the ABEP of convolutional coded $2\times 1$ and $2\times 1$ OSM systems are depicted, assuming similar propagation conditions to those considered in Fig.~\ref{Fig:ABEP2}. Considering a convolutional code with rate $1/3$ and constraint
length of 3, its transfer function is given as \cite[Eq. (8.2.6)]{B:Proakis}
\begin{equation}\label{Eq:Tfunction}
T[D(\theta), N] = \frac{D(\theta)^6 N}{1-2ND(\theta)^2}.
\end{equation}
Substituting \eqref{Eq:Tfunction} to \eqref{Eq:bound2}, a union bound on the ABEP can be obtained as
\begin{equation}\label{Eq:bound3}
\bar{P}_{\mathrm{ub}} \leq \frac{1}{\pi\log_2(M)}\int_0^{\pi/2}\frac{D(\theta)^6}{(1-2D(\theta)^2)^2}\mathrm{d}\theta.
\end{equation}
The performance results of Fig.~\ref{Fig:Coded} clearly show that, as expected, the incorporation of convolutional coding significantly enhances the performance of OSM systems, even when a small number of receive apertures is employed, even for $N = 1$.

\section{Conclusion}\label{Sec:Conclusions}

In this paper, the use of spatial modulation technique for coherent FSO communication systems has been proposed. We have provided a comprehensive analytical framework for error performance
analysis in the presence of atmospheric turbulence scattering channel models which include the H-K distribution.
The proposed framework reveals important information about the performance of OSM over such turbulent channels, including the effect of fading
severity and the achievable diversity gain. It also provides valuable insight into the impact of channel parameters on
performance of OSM.
Upper bounds for the ABEP performance of coded OSM systems have also been derived, demonstrating that
coding techniques can greatly enhance the performance of OSM.
Extensive computer simulation performance evaluation results have been also obtained which have verified the accuracy of the analytical approach.
Important trends about the performance of OSM for a variety of atmospheric turbulent scenarios and MIMO setups have also been identified. For example, it was shown that OSM can provide significant performance enhancements in the presence of atmospheric turbulence. The improvements are comparable to the ones offered by conventional coherent systems with spatial diversity, while outperforming the latter in terms of spectral efficiency and hardware complexity. Besides, under specific propagation conditions, OSM can yield better performance than conventional SIMO systems employing MRC or SC.
We believe that the proposed framework is a useful tool for understanding the performance trend, important properties and tradeoffs of outdoor OSM operating in the presence of atmospheric turbulence.
\bibliographystyle{IEEEtran}

\end{document}